\documentclass{aa}
\usepackage{amssymb}
\usepackage{amsmath}
\usepackage{xspace}
\usepackage{graphicx}

\usepackage{natbib}

\def\gr{$\gamma$-ray}
\def\fermi{Fermi\xspace}

\authorrunning{Neronov \& Malyshev}

\begin{document}
\title{Hard spectrum of cosmic rays in the  Disks of Milky Way and Large Magellanic Cloud.}
\author{A. Neronov $^{1}$, D. Malyshev $^{1}$}
\institute{
1. ISDC, Astronomy Department, University of Geneva, Ch. d'Ecogia 16, 1290, Versoix, Switzerland \\
}

\abstract
{The slope of the locally measured spectrum of cosmic rays varies from  2.8 for protons with energies below 200~GeV down to 2.5 for heavy nuclei with energies in the TeV-PeV range. It is not clear if the locally measured slope values are representative for those of the overall population of Galactic cosmic rays and if the slope of the cosmic ray spectrum varies across the Galaxy, e.g. in response to the variations of the star formation rate.  }
{We use the data of Fermi Space \gr\ Telescope to derive a measurement of the slope of the cosmic ray spectrum across the Galactic Disk and to compare it with that of the Large Magellanic Cloud cosmic rays. }
{ A special choice of the background estimation regions allows us to single out the neutral pion decay component of the  \gr\ flux in the energy range above 10~GeV and to separate it from (a) emission from the local interstellar medium around the Solar system and (b) from the inverse Compton emission produced by cosmic ray electrons. } 
{The spectrum of the pion decay \gr\ emission from the Galactic disk  in the energy band 10~GeV -- 1~TeV has the slope $\simeq 2.4$. There is no evidence for the variation of the slope with Galactic longitude / distance from the Galactic Centre.    
The slope of the spectrum of cosmic rays derived from the \gr\ data, $\simeq 2.45$,  is harder than the slope of the locally observed cosmic ray proton spectrum. Pion decay emission from a powerlaw distribution of cosmic rays with the same hard slope also provides a fit to the \gr\ spectrum of the Large Magellanic Cloud. }
{ Identical and hard slopes of the spectra of cosmic rays in the Milky Way and in the Large Magellanic Cloud are consistent with a straightforward theoretical model in which cosmic rays are injected by shock acceleration with the spectrum with the slope $\Gamma\simeq 2 ... 2.1$ which is subsequently modified by $\Delta\Gamma=1/3 ... 1/2$ by the energy-dependent escape of cosmic rays through the turbulent Galactic magnetic field. Deviations of the locally measured cosmic ray spectrum from the average Galactic spectrum are explained by the discrete distribution of the cosmic ray sources in space and time.  } 

\keywords{Gamma rays; Star forming regions}

\maketitle

\section{Introduction}

Our knowledge of the properties of Galactic cosmic rays (CRs) is based mostly on the measurements in a single point in the Milky Way galaxy: the position of the Solar system. From these local measurements we know that the spectrum of CRs is an approximate powerlaw with the slope ranging from 2.8 below $200$~GeV to 2.6 ... 2.7 in the TeV-PeV energy range, ending with the "knee" feature above the PeV energy \citep{pdg}. 

Numerous theoretical models were developed over time to explain the spectral properties of the cosmic ray flux \citep{berezinsky_book,blasi_review}. An implicit assumption of these models is that the approximate 2.7 slope powerlaw spectrum is "universal", in the sense that it is valid all over the Milky Way galaxy. The same assumption is commonly adopted in the interpretation of the data on \gr\ emission from Galactic CR population \citep{fermi_diffuse_2012} and in the modelling of the extragalactic \gr\ background produced by a population of star-forming galaxies in the Universe \citep{fermi_starforming}.

An assumption that the Galactic CR spectrum has the slope 2.7 everywhere across the Galaxy is, however, far from being obvious. None of the theoretical models which "explains" the spectral slope is developed from the first principles. The match between the observed spectral slope and the model calculations is achieved via adjustment of phenomenological parameters, e.g. of the slope of the injection spectrum of CRs in the interstellar medium, $\Gamma_{inj}$ and the modification of the slope by the energy dependent escape of the CRs, characterised by the energy slope of the diffusion coefficient, $\delta$ \citep{berezinsky_book}. It is well possible that the slope of the locally observed CR spectrum is not identical to that of the average Galactic CR spectrum. Variations of the star formation rate over the last tens of millions of years and discreteness of the cosmic ray source distribution could lead to fluctuations of the shape of the CR spectrum in space and time \citep{aharonian_book,TeV_sources,kachelriess}. 

The only possibility for the measurement of the spectrum of CRs at different locations in the Galaxy is via the use of the \gr\ data  \citep{aharonian_book,fermi_diffuse_2012,neronov_GouldBelt}. CRs interacting in the interstellar medium (ISM) produce neutral pions which decay into photons. The spectrum of the pion decay \gr s has the slope which is nearly identical to that of the parent CR spectrum. 

An immediate problem for the measurement of the CR spectrum from the \gr\ data is that the \gr\ flux from the Galaxy contains, apart from the pion decay flux, also the flux produced by the CR electrons via the Bremsstrahlung and inverse Compton mechanisms \citep{fermi_diffuse_2012}. As a consequence, derivation of the CR spectrum from the \gr\ data should include a detailed modelling of the three contributions, or a method of separation of the pion decay contribution to the flux from the Bremsstrahlung and inverse Compton contributions. 

In what follows we develop a method which allows to separate the pion decay emission from the Bremsstrahlung and inverse Compton emission in the flux coming from the distant parts of the Disk of the Milky Way galaxy. Using this method, we are able to measure the spectrum of cosmic rays at different locations in the Disk, at different distances from the Centre of the Galaxy. We show that the average Galactic CR spectrum  in the TeV energy band is harder than the locally observed spectrum of the proton component of the CR flux. We also demonstrate that the average Galactic CR spectrum slope does not vary significantly across the Galaxy and it is also identical to the slope of CRs filling the disk of the Large Magellanic Cloud (LMC) galaxy. This suggests that the measured average slope of the CR spectrum is "universal", in the sense that it is determined by the basic physical laws. It is this slope which could be explained by the straightforward theoretical models of of CR injection and propagation.

\section{Data Analysis}
\label{sec:Data_Analysis}

In our work we use 6.7~years of \fermi/LAT data (from August, 4th, 2008 to December 2014). We process the data using the most recent version of the \fermi Science Tools Software v9r33p0 and the P7REP response functions\footnote{http://fermi.gsfc.nasa.gov/ssc/data/analysis/}. The spectra of the sources are calculated using two complementary techniques: the binned likelihood analysis and the aperture photometry. 

The binned likelihood analysis is based on the fitting of a model of diffuse and point source emission within a "region of interest"  to the data. The spatial extent of the region is $13^\circ$ radius at energies $\gtrsim 300$~MeV and $20^\circ$ radius in the 60-300~MeV energy range.   The spatial model includes diffuse Galactic and extragalactic backgrounds and the sources from the 4 year (3FGL) \fermi catalog \citep{fermi_catalog}. In each  energy bin we fix the spectral shape of each source to be a power law with index $-2$. The spectral shapes of diffuse backgrounds are given by the corresponding templates. The normalisations of fluxes of all sources and diffuse backgrounds are treated as free parameters during the fitting.

The aperture photometry method of spectral analysis uses a selection of \gr\ events from pre-defined  signal and background regions with the {\it gtbin / gtmktime} tools and the calculation of the exposure for the selected regions with the {\it gtexposure} tool. All the sources have the spatial extent much larger than the size of the Point Spread Function (PSF) of LAT in the energy range above 1~GeV, where the aperture photometry analysis is performed. Taking this into account, we use the {\it gtexposure} tool with the switch {\tt appcor=NO}, which precludes the correction of exposure for the PSF effects. We remove the contribution of bright point sources to both the signal and background regions by excluding circles of the radius $0.5^\circ$ around identified pulsars and active galactic nuclei from the Fermi catalog \citep{fermi_catalog} and around the Galactic Centre source. In the analysis of the signal from the Galactic Plane we explicitly do not remove the unidentified sources from the catalog, because a large fraction of these sources could be just the inhomogeneities of the the overall diffuse emission from the ISM. We also do not remove the supernova remnants because they could be considered as a part of the overall \gr\ signal from the interacting Galactic CRs. 

\section{The Galactic Disk}

The  Galactic diffuse emission is a sum of the pion decay, Bremsstrahlung and inverse Compton contributions. Relative importance of different contributions to the \gr\ flux depends on the photon energy and on the location of the emission region in the Galaxy. 

Our goal is to derive a measurement of the spectrum of proton / nuclei CRs residing at different locations in the Galactic Disk from the spectrum of the pion decay emission from the Disk. There are two obvious obstacles for such a measurement. First, the  pion decay flux has to be separated from the Bremsstrahlung and inverse Compton fluxes. Next, the flux of the pion decay emission produced at large, several kiloparsecs distances from the Sun has to be separated from the flux of pion decay and inverse Compton / Bremsstrahlung emission form the local interstellar medium. 

The spectrum of the Bremsstrahlung is soft and its contribution to the \gr\ flux is negligible in the energy band above 10~GeV \citep{fermi_diffuse_2012}. Thus, reducing the energy range of the analysis to $E_\gamma>10$~GeV should provide a signal which is largely free from the Bremsstrahlung contribution.

The spectrum of inverse Compton emission is harder and the inverse Compton flux could be significant up to the highest energies accessible for the Fermi/LAT. The inverse Compton flux is strongest close to the direction of the Galactic Plane and toward the inner Galaxy, because of the higher density of the interstellar radiation field in the inner part of the Galactic Disk. 

Fortunately, the pion decay \gr\ emission is still more concentrated toward the Galactic Plane because its volume luminosity scales with the density of the interstellar gas. The luminosity is falling exponentially with the increase of the altitude above / below the Galactic Plane because the density of the Galactic Disk is falling exponentially with the scale height about 100~pc. 

\begin{figure*}
\includegraphics[width=0.55\linewidth]{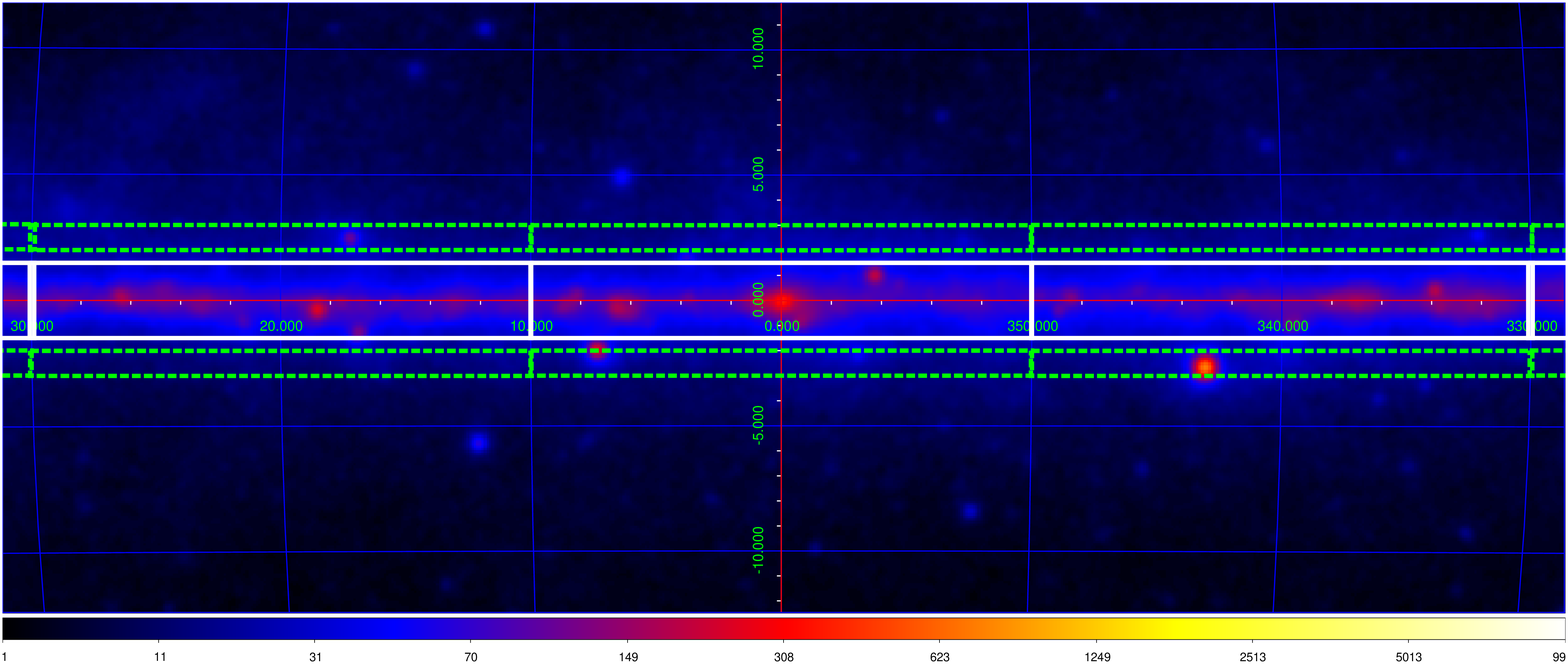}
\includegraphics[width=0.44\linewidth]{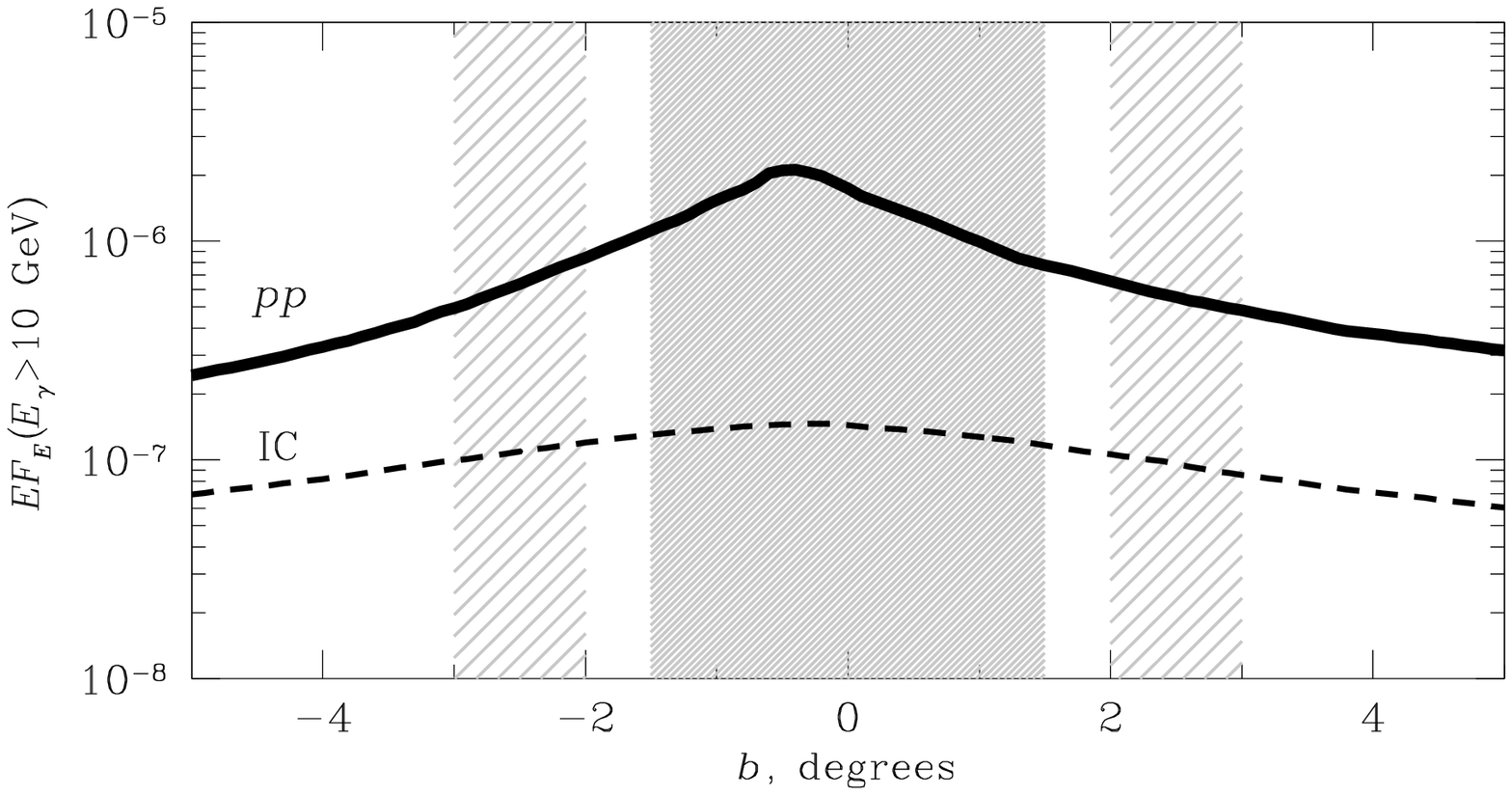}
\caption{Left: Fermi/LAT count map of the inner Galaxy region above 1~GeV. White solid boxes show the $3^\circ\times 20^\circ$ regions of signal extraction, dashed green boxes show the regions of background estimate.  Right: Galactic latitude profiles of the \gr\ emission from $pp$ interactions and of the inverse Compton emission, from \citet{fermi_diffuse_2012}. Dense and light shading show the ranges of Galactic latitudes from which the signal and background are extracted for the calculation of the spectrum.  }
\label{fig:profile_b}
\end{figure*}

This difference leads to a significant difference of the Galactic latitude profiles of the pion decay and inverse Compton emission \citep{fermi_diffuse_2012}, see Fig. \ref{fig:profile_b}, right panel. To separate the pion decay and inverse Compton signals,  we take the source signal from a narrow strip $|b|<1.5^\circ$ around the Galactic Plane and estimate the background from two narrow strips $2^\circ<|b|<3^\circ$ as it is shown in the left panel of Fig. \ref{fig:profile_b}. The pion decay emission profile is sharply peaked toward $b=0^\circ$, compared to the inverse Compton profile. Integrating the pion decay and inverse Compton fluxes in the $|b|<1.5^\circ$ and $2^\circ<|b|<3^\circ$ regions one could find that the choice of the source and background regions shown in the left panel of Fig. \ref{fig:profile_b} rejects $\simeq 80\%$ of the inverse Compton signal within $|b|<1.5^\circ$ strip, while leaving $>50$\% of the pion decay flux. In this way the pion decay flux, which (within the model of \citet{fermi_diffuse_2012}) was already an order of magnitude higher than the inverse Compton flux within the signal extraction strip, gets boosted by another half-an-order. 

The same choice of the signal and background regions allows also the suppression of the flux from the local ISM. The angular extent of the source occupying a 100~pc wide layer around the Galactic Plane is about $10^\circ$ at the distance 1~kpc from the Sun and is just about $3^\circ$ for the flux coming from the distances beyond 3~kpc from the Sun position. The highest star formation rate in the Milky Way is in the innermost part of the Galactic Disk at the distances $\gtrsim 4$~kpc from the Sun. This innermost part of the Galactic Disk spans a strip $|b|<3^\circ$ on the sky. Choosing both the signal and background regions within a region where the flux from the local ISM does not vary removes the local ISM contribution.

\begin{figure}
\includegraphics[width=\linewidth]{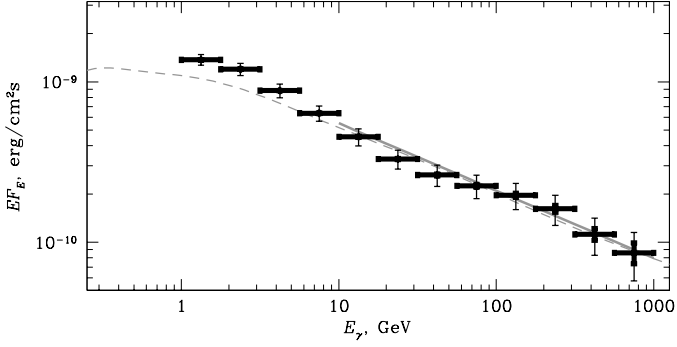}
\caption{Spectrum of emission from the $|l|<90^\circ$, $|b|<1.5^\circ$ part of the Galactic Plane. Thick / thin errorbars show the statistical / systematic error. Grey thick line shows the best-fit powerlaw with the slope $\Gamma_\gamma=2.42$ in the energy range above 10~GeV. Dashed line shows the spectrum of the neutral pion decay emission produced a powerlaw proton spectrum with the slope $\Gamma_p= 2.45$.   }
\label{fig:spectrum_GalPlane}
\end{figure}

The spectrum of diffuse emission from $|b|<1.5^\circ,\ |l|<90^\circ$ strip extracted using the aperture photometry method is shown in Fig. 
\ref{fig:spectrum_GalPlane}.  The spectrum above 10~GeV could be fit by a powerlaw with the slope $\Gamma_\gamma=2.42\pm 0.03_{stat}\pm 0.12_{syst}$. In this energy band the flux is dominated by the pion decay emission. The spectrum could be fitted with a model pion decay spectrum calculated using the code of \cite{kamae} for the powerlaw proton spectrum with the slope $\Gamma_p=2.45$. 

Below 10~GeV the \gr\ spectrum gets softer. This softening  is expected at least  due to a non-negligible contribution of the electron Bremsstrahlung to the photon flux. Alternatively, the pion decay spectrum itself could soften below 10~GeV is the CR spectrum gets softer below $\sim 150$~GeV, as does the local CR spectrum. Our analysis method does not allow to distinguish between these two possibilities. 

\begin{figure}
\includegraphics[width=\linewidth]{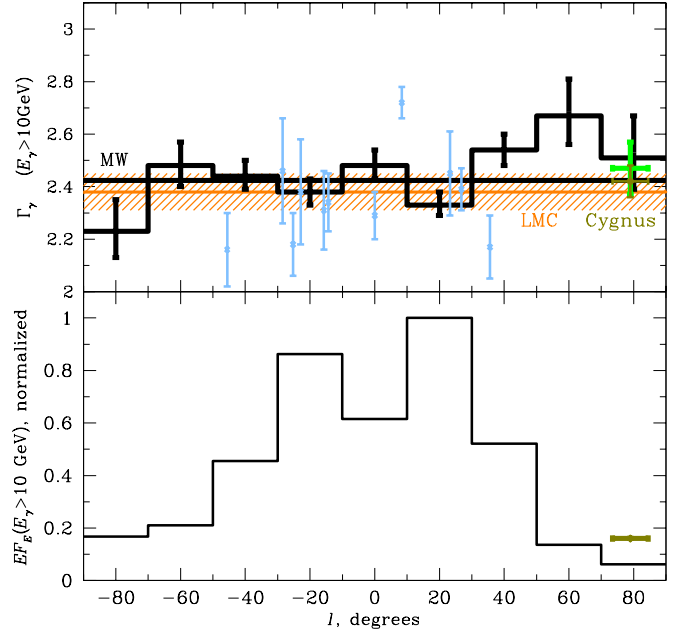}
\caption{Galactic longitude profile of the flux (bottom panel) and slope (top panel) of \gr\ emission from the Galactic Disk in the energy band above 10~GeV.  Horizontal black line shows the best fit spectral slope.  Orange lthin line and the hatched region show the slope of the \gr\ spectrum of the LMC. Green and olive data points at $l\simeq 80^\circ$ show the slope of the spectrum of Cygnus region extracted using the likelihood and aperture photometry methods.  Light blue data points show the slopes of the spectra of unidentified sources from the HESS Galactic Plane survey. }
\label{fig:profile_gamma}
\end{figure}

It is not clear a-priori if the average slope of the CR spectrum depends on the distance from the centre of the Galaxy as suggested by \cite{gaggero} or it has a "universal" slope. It is also not clear if the slope depends on the relevant physical parameters like e.g. the star formation rate. Fig. \ref{fig:profile_gamma} shows the slope of the \gr\ spectrum of the $|b|<1.5^\circ$ strip as a function of Galactic longitude. One could see that there is no evidence for the dependence of the slope on the Galactic longitude or, equivalently, on the distance from the Galactic Centre. There is also no correlation between the slope and the overall flux above 10~GeV, as one could see from comparison of the upper and lower panels of Fig.  \ref{fig:profile_gamma}.

\subsection{Cygnus region}

The method of calculation of the pion decay contribution to the diffuse flux from the Galactic Disk developed in the previous section does not work in the outer Galaxy, in the Galactic longitude range $|l|\gtrsim 90^\circ$. This is because the contribution of the distant parts of the Galactic Disk at several kiloparsecs distances becomes much smaller than the flux from the local ISM. In this case, estimation of the flux from a narrow strip around $b=0^\circ$ plane catches the degree-scale  fluctuations of the flux from the local Galaxy, rather than the flux from the distant parts of the Galactic Disk. 

A measurement of the spectrum of CRs in a distant part of the Galactic Disk is still possible in one particular case, namely in the direction of Cygnus region. This direction is tangent to the Galactic arm passing close to the Sun. An enhancement of the diffuse emission from the direction of Cygnus region is commonly attributed to a superposition of the projected \gr\ emission from the local Galactic arm and the flux produced by the nearby active star formation region at $\sim 1.5$~kpc distance in this direction.  Emission from this nearby star forming region is known to have a hard spectrum with the slope close to $\Gamma_\gamma\simeq 2.2$, supposedly due to a recent injection of freshly accelerated CRs \citep{fermi_cocoon}.

\begin{figure}
\includegraphics[width=\linewidth]{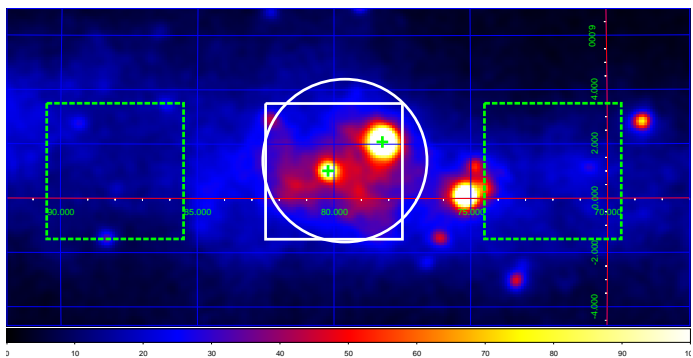}
\caption{Fermi/LAT count map of Cygnus region in the energy band above 1 GeV. White circle shows the extent of the Gaussian model template for the spectral extraction. White solid and green dashed boxes show the regions of signal and background extraction for the calculation of the spectrum using the aperture photometry. }
\label{fig:cygreg}
\end{figure}

\begin{figure}
\includegraphics[width=0.45\textwidth]{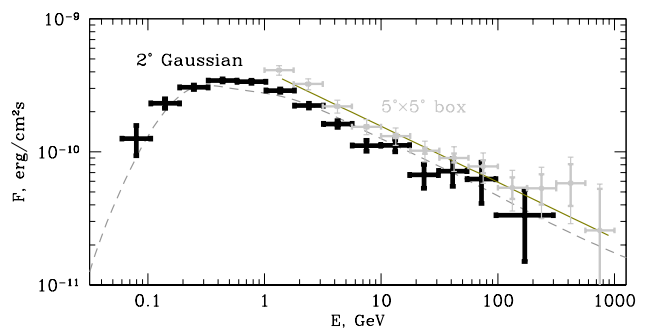}
\caption{Spectrum of the Cygnus region extracted from a $2^\circ$ Gaussian spatial template (thick data points) and from an $5^\circ\times 5^\circ$ box. Grey dashed line shows a fit with a model of pion decay emission from the powerlaw CR spectrum with the slope $\Gamma_{CR}=2.45$. Solid grey line shows a  powerlaw fit to the box spectrum. }
\label{fig:spectrum_Cygnus}
\end{figure}

The extended emission source in the Cygnus region, reported in the Fermi catalog, is described by a Gaussian spatial template with the width $2^\circ$, shown in Fig. \ref{fig:cygreg}. Spectrum of the source with such spatial morphology is shown in Fig. \ref{fig:spectrum_Cygnus}. The spectrum above 3 GeV is well described by a powerlaw with the slope $\Gamma_\gamma=2.47\pm 0.09$. This slope is softer than the slope reported by \cite{fermi_cocoon}, because we have used, contrary to  \cite{fermi_cocoon}, the standard Galactic diffuse emission background for the spectral modelling. In such standard model the extended source in the direction of Cygnus includes the entire emission from the Galactic arm behind the nearby star formation region, rather than only a contribution from the nearby star formation region at 1.5~kpc distance.  

As a cross-check, we extract the spectrum of the Cygnus region using the aperture photometry method and taking the source region to be a square box of the size $5^\circ\times 5^\circ$ centered at $l=80^\circ, \ b=1^\circ$, see Fig. \ref{fig:cygreg}.  Similarly to the Galactic Plane spectrum case, we suppress the contribution to the flux from the local interstellar medium and from the inverse Compton emission by an appropriate choice of the background estimate regions. Both the local interstellar medium and inverse Compton contribution do not vary significantly along the Galactic Plane in the Galactic longitude range $70^\circ<l<90^\circ$. Choosing the boxes of the same size as the signal box, but displaced toward smaller and larger $l$, as it is shown in Fig. \ref{fig:cygreg} suppresses the two components of the diffuse emission, leaving the pion decay and Bremsstrahlung flux from the Cygnus region unaffected. 

The spectrum of emission from the $5^\circ\times 5^\circ$ box is shown in Fig. \ref{fig:spectrum_Cygnus} by the thin olive colour data points. The slope of the spectrum is $\Gamma_\gamma=2.42\pm 0.06$, which is consistent with the slope of the spectrum extracted using the Gaussian template. 

The emission from the Cygnus region is dominated by the pion decay flux, as it is clear from the presence of the low-energy cut-off in the spectrum at the pion production threshold $E_\gamma\simeq 100$~MeV \citep{pion_bump}. The slope of the pion decay spectrum in the Cygnus region is identical to the slope of the spectrum elsewhere in the Galactic Plane. This confirms that the slope $\Gamma_\gamma\simeq 2.42$ is characteristic for the slope of the CR spectrum produced by star formation activity and that it is largely independent of the parameters of the interstellar medium. Indeed, the properties of the interstellar medium in Cygnus region and in the local Galactic arm behind the Cygnus region are quite different from those in the innermost arms and in the Galactic Bar. However, the slopes of the CR spectra in these regions are the same. 

\section{Large Magellanic Cloud}

The conjecture on the universality of the spectrum of CRs produced by the star formation could be tested via observations of other galaxies. The closest to us disk galaxy is the Large Magellanic Cloud (LMC) at the distance 50~kpc \citep{lmc_distance}. 

\begin{figure}
\includegraphics[width=\linewidth]{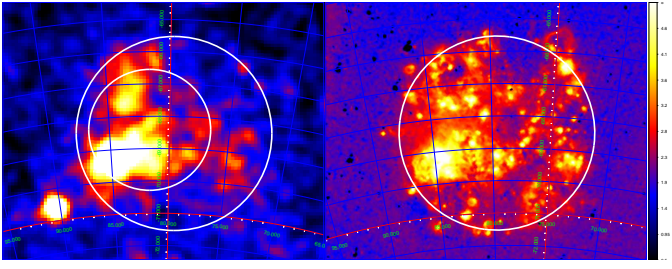}
\caption{Fermi/LAT count map of LMC field in the energy band above 1~GeV (left) compared to the H$\alpha$ map of LMC (right). Smaller white circle in the left panel shows the extent of the Fermi/LAT Gaussian template for the spatial model of LMC. Larger circle shows the extent of the source signal extraction region for the aperture photometry calculation of the spectrum.}
\label{fig:LMC}
\end{figure}

\begin{figure}
\includegraphics[width=\linewidth]{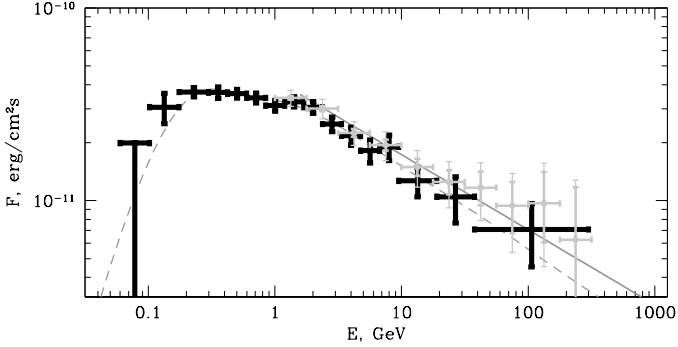}
\caption{Spectra of the LMC extracted using the likelihood analysis (thick data points) and the aperture photometry method (grey thick / thin data points showing the statistical / systematic errors). Dashed line shows a model fit with the pion decay spectrum produced by the parent proton population with the powerlaw spectrum with the slope $\Gamma_{CR}=2.45$. Solid line shows a powerlaw fit to the aperture photometry spectrum.}
\label{fig:LMC_spectra}
\end{figure}

The Fermi/LAT count map of the LMC field is shown in the left panel of Fig.~\ref{fig:LMC}.  The brightest emission is detected from the direction of 30 Doradus star forming region, which is also visible as a bright excess on the H$\alpha$ map of the LMC shown in the right panel of Fig. \ref{fig:LMC} \citep{Halpha}. The overall extent of the LMC disk is at least $3^\circ$, as it is clear from the H$\alpha$ map. However, the \gr\ brightest part of the LMC is more compact, fitting within a circle of  $\simeq 1.9^\circ$ in radius. Apart form the 30 Doradus region, significant flux comes from the Constellation III star forming region North of the 30 Doradus.    

The overall spatial model of the \gr\ emission from LMC, considered by \citet{FERMI_LMC} is a sum of two gaussians of $\sim 1.9^\circ$ and  $\sim 0.2^\circ$ widths. The smaller size Gaussian is centred at the position of the  30 Doradus  region. The spectrum of LMC extracted using this template is shown in Fig. \ref{fig:LMC_spectra}.  Above 3~GeV, the LMC spectrum is well described with a power law model with the index $\Gamma = -2.41 \pm 0.15$. 

Similarly to the case of the Cygnus region, we cross-check the result using the aperture photometry spectral extraction method. We take the signal region to be a circle of the radius $3^\circ$, centered at RA=79.61$^\circ$, DEC=-68.53$^\circ$, which corresponds to the position and extent of the H$\alpha$ disk of the LMC, see Fig. \ref{fig:LMC}. We estimate the background from a circle of equal radius centered RA=70.82$^\circ$, DEC=-76.75$^\circ$,  situated at the same Galactic latitude as the source circle.  The spectrum extracted using the aperture photometry method is shown by thin data points in Fig. \ref{fig:LMC_spectra}. It could be fit by a powerlaw with the slope $\Gamma_\gamma=2.39\pm 0.09_{stat}\pm 0.07_{syst}$, which is consistent with the slope of the spectrum extracted using the likelihood analysis.

The overall spectrum in the broad energy range from 60~MeV up to 100~GeV is well fit with a single pion decay component spectrum with the slope of the parent CR spectrum $\Gamma_{CR}=2.45$, shown by the dashed line in Fig. \ref{fig:LMC_spectra}. Contribution of the electron Bremsstrahlung component to the spectrum is constrained by the upper limit in the 60-100~MeV band, where the Bremsstrahlung flux should dominate over the pion decay flux. Similarly to the Galactic Plane and Cygnus region spectra, the Bremsstrahlung flux in the energy range above several GeV is negligibly small. 

The slope of the \gr\ spectrum of the LMC above 3~GeV is consistent with the slope of the spectrum of \gr\ emission from the Milky Way disk, see Fig. \ref{fig:profile_gamma}. This suggests that the spectrum of emission from the LMC disk is dominated by the pion decay flux from the proton / nuclei CR interactions in the LMC disk, similarly to the Milky Way emission in the same energy band. This conclusion is supported by the analysis of \citet{foreman15} who find that the pion decay flux dominates the \gr\ emission from the LMC except possibly for the 30 Doradus region where electron Bremsstrahlung and inverse Compton emission could compete with the pion decay emission.

\section{Other star forming galaxies}

Apart from the Milky Way and LMC, several other normal star forming Galaxies are detected by Fermi/LAT. The conjecture of the universal hard slope of the CR spectrum produced by the star formation should, in principle, hold also for these galaxies. 

We have extracted the spectra of the Small Magellanic Cloud and M31  galaxies and checked that the quality of the spectra is not yet sufficient for the measurement of the details of the spectral properties of the CR populations in these galaxies.

\section{Discussion}

Our analysis of the spectral properties of \gr\ signal from a $3^\circ$ wide strip along the Galactic Plane has enabled a separation of the pion decay emission component of the Galactic diffuse emission from the Bremsstrahlung and inverse Compton components in the energy band above 10~GeV. We have used the measurement of the slope of the pion decay spectrum to measure the slope of the spectrum of CRs with energies above 100~GeV across the Galactic Disk. 

The slope of the Galactic cosmic ray spectrum turns out to be harder than that of the locally measured CR spectrum in the same energy band. The slope of the Galactic CR spectrum does not exhibit significant variations from the inner to outer Galaxy. This points to a "universal" nature of the hard slope, which is largely independent of the local physical conditions in different parts of the Galaxy. 

The same hard slope of the CR spectrum is found also in the LMC galaxy. This further supports the conjecture about the "universality" of the hard slope of the spectra of CRs resulting from star formation in galaxies. 

The slope of the average CR spectrum derived from our analysis of the Galactic Plane and of the LMC coincides with the slope of the neutrino spectrum in the energy band above 10~TeV, reported by the IceCube collaboration \citep{icecube_slope}. This suggests that the astrophysical neutrino flux has a contribution from the Milky Way  \citep{neronov_GalNeutrino1,neronov_GalNeutrino}. A smooth match between the diffuse Galactic \gr\ emission spectrum in the sub-TeV energy band with the IceCube neutrino spectrum extending up to the PeV energy range indicate that the hard powerlaw CR spectrum with the slope $\Gamma_{CR}\simeq 2.45$ extends over several decades in energy, from $\lesssim 1$~TeV up to $\sim 10$~PeV. 

In the 1-10~TeV energy range, not covered by the Fermi/LAT and IceCube data, the data of the ground-based \gr\ telescopes could provide a measurement of the characteristic slope of the Galactic CR spectrum. Such a measurement is more challenging because of the lack of the wide-field telescopes in this energy band. However, an assessment of the average CR spectrum in the Galactic Disk is still possible using the narrow field-of-view telescopes like e.g. HESS. This is illustrated in Fig. \ref{fig:profile_gamma}. The thin light blue data points in the upper panel of this figure show the measurements of the slopes of the spectra of unidentified extended sources in the Galactic Plane, from the HESS Galactic Plane survey  \citep{HESS_survey0,HESS_survey,HESS_survey1}. Fairly tight correlation of the positions of the extended HESS survey sources with the still more extended sources detected above 100~GeV in the Fermi/LAT data \citep{TeV_sources} indicates that the HESS unidentified sources are parts of larger excesses of the diffuse emission occurring  at the positions of over-densities in the interstellar medium and/or close to the most recent points of injection of CRs in the ISM. If this is so, a spread in the slopes of the \gr\ spectra of these sources is expected because the dense clumps of the ISM could be situated closer or further from the recent points of injection of CRs \citep{aharonian_book,kachelriess}. At the same time, the average-over-the-sources slope of the CR / \gr\ spectra should be close to the average slope of the spectrum of Galactic CRs. 
This is indeed the case. The average slope of the spectra of sources from the HESS survey is remarkably close to the slope of the spectrum of the $|b|<1.5^\circ$ strip, $\Gamma_{HESS}\simeq 2.43$. 

The value $\Gamma_{CR}\simeq 2.45$ of the universal slope of the spectrum  of CRs produced by the star formation process could be understood within a straightforward theoretical model, without adjustment of phenomenological parameters. The most commonly considered process of the CR production is the shock acceleration process. This process results in the slope of the injection spectrum of CRs $\Gamma_{inj}\simeq 2  ... 2.1$ in the most simple settings \citep{berezinsky_book,blandford,berezhko,malkov,bell01,blasi_review}. Propagation of CRs through the turbulent Galactic magnetic field with the subsequent escape of CRs from the Galaxy leads to the softening of the CR spectrum from $\Gamma_{inj}$ to $\Gamma_{CR}=\Gamma_{inj}+\delta$ where $\delta$ is related to the slope of the turbulence power spectrum. It is $\delta=1/3$ for the Kolmogorov turbulence and $\delta=1/2$ for the Iroshnikov-Kraichnan turbulence. Measurement $\Gamma_{CR}\simeq 2.45$ suggests that the injection spectrum of CRs in the ISM has the slope $\Gamma_{inj}=2.1$, assuming the Kolmogorov spectrum of the ISM turbulence \citep{ISM_turbulence}.

Also straightforward is the understanding of the difference of the slopes of the average Galactic CR spectrum and the CR spectra measurable at fixed locations in the Galactic Disk, like e.g. the measurements at the position of the Solar system.  The spectra of CRs measured in any single point typically deviate from the average CR spectrum because they are influenced by the discreteness of the CR source distribution in space and time. A single source which injects about $10^{50}$~erg in CRs  could significantly alter the overall energy density of CRs  in the local ISM within the region of the volume more than $(0.1 \mbox{ kpc})^3$ on the time scales of millions of years \citep{aharonian_book,cream,kachelriess}. In a similar way, a star formation activity episode resulting in formation of a superbubble with an energy output $10^{52}$~erg would change the CR spectrum in a kpc-scale region around it for tens of millions of years \citep{neronov_GouldBelt}. In general, the CR spectra measurable at locations  close to recently active sources are harder than the  CR spectra  in a places where no recent injections of CRs happened on the time span comparable to the escape time of CRs from the Galactic Disk \citep{aharonian_book,TeV_sources,kachelriess}. The slopes of the CR spectra shown in Figs.  \ref{fig:spectrum_GalPlane}, \ref{fig:profile_gamma}, \ref{fig:spectrum_Cygnus}, \ref{fig:LMC_spectra} are the same only because they are derived from the \gr\ signal collected from sufficiently large regions of the ISM, where the assumptions on the continuity and time independence of the source distribution are valid. 

\section{Summary}

We have used the \gr\ data to derive a measurement of the characteristic spectrum of CRs produced by the star formation activity in the Milky Way (Fig. \ref{fig:profile_gamma}) and in the LMC (Fig. \ref{fig:LMC}) galaxies. The slope of the average spectra of CRs in these two galaxies, $\Gamma_{CR}\simeq 2.45$,  is harder than the locally measured spectrum of CRs penetrating in the Solar system. Consistency of the measurement of the slopes of the CR spectra at different locations in the Milky Way Galactic Disk and in the LMC suggests that the spectrum of CRs produced by the star formation activity is universal, i.e. it does not depend on the details of the physical parameters of the star formation process and of the ISM. The value of the universal slope of the CR spectrum  is consistent with the straightforward theoretical model of CR injection by the shock acceleration process, which produces the spectrum with the slope $\Gamma_{inj}\simeq 2 ... 2.1$, followed by the energy-dependent escape of CRs through the turbulent Galactic magnetic field with the ISM turbulence power spectrum. Fluctuations of the slopes of the CR spectra measurable at fixed points in the ISM are attributed to the discreteness of the CR source distributions and variations of the star formation rate in time. 

\bibliography{bib}
\end{document}